
\magnification =\magstephalf
\hsize=5.25 in
\hoffset=1 true cm\voffset=1.5 true cm 
\vsize=7 in
\baselineskip=12pt
\parskip=0.5\baselineskip
\parindent=2.5em
\font\bigbold=cmbx12 scaled\magstephalf

\font\medcaps=cmcsc10 scaled\magstephalf
\font\medslant=cmsl10 scaled\magstephalf

\font\smallbold=cmbx9
\font\smallital=cmti9
\font\smallroman=cmr9

\def\daectitle#1{
        \vfill\eject
        \centerline{
                \bigbold
                #1
        }
}

\def\daecauthors#1{
        \vskip 1.5\baselineskip
        \centerline{
                \medcaps
                #1
        }
}

\def\daecinst#1{
        \vskip 0.9\baselineskip
        \centerline{
                \medslant
                #1
        }
}
\def\daecinsttwo#1{
        \vskip 0.5\baselineskip
        \centerline{
                \medslant
                #1
        }
}
\def\daecinstthree#1{
        \vskip 0.5\baselineskip
        \centerline{
                \medslant
                #1
        }
}
\def\daecabstract{
        \vskip\baselineskip
        \centerline{\bf Abstract}
        \noindent
}
\def\daecsec#1{
        {
                \parindent=0pt
                \vskip\baselineskip
                \vbox{
                        \bf #1
                        \vskip -0.2\baselineskip
                }
        }
}

\def\daecakno{
        \par
        \noindent{\bf Acknowledgements:\ }
}

\def\daecfigure#1#2#3{
        \goodbreak
        \midinsert
                \leftskip=1 true cm
                \rightskip=1 true cm
                \vskip#2 true cm
                \noindent{\bf Figure #1.} #3
                \vskip-0.5\baselineskip
                \leftskip=0 true cm
                \rightskip=0 true cm
        \endinsert
}
\newcount\figureno
\def\daecfiguretwo#1#2#3#4{
        \goodbreak
        \midinsert
                \vskip#2 true cm
                \noindent{\bf Figure #1.} #3
                \par
                \figureno=#1
                \advance\figureno by 1
                \noindent{\bf Figure \number\figureno.} #4
                \vskip-0.5\baselineskip
        \endinsert
}
\def\daecrefhead{
        \vskip\baselineskip
        \parindent=0pt
        \parskip=0pt
        \vbox{
                \centerline{\bf References}
                \vskip0.5\baselineskip
        }
}
\def\daecpaper#1#2#3#4#5{
        \hangindent=3em
        \hangafter=1
        \smallroman #1, #2,
        {\smallital #3\/},
        {\smallbold #4}, #5.
        \par
}

\def\daecpreprint#1#2{
        \hangindent=3em
        \hangafter=1
        \smallroman #1, #2, preprint.
        \par
}

\def\daecbook#1#2#3#4#5{
        \hangindent=3em
        \hangafter=1
        \smallroman #1, #2,
        {\smallital #3\/} (#4: #5).
        \par
}

\def\daecproceed#1#2#3#4#5#6#7{
        \hangindent=3em
        \hangafter=1
        \smallroman #1, #2,
        {\smallital in #3\/},
        ed. #4 (#5: #6), p. #7.
        \par
}

\def\apj{Ap. J.}

\def\aa{Astr. Ap.}

\def\mnras{M.N.R.A.S.}

\def\nature{Nature}

\def\etal{{\it et al.\/}\ }

\tabskip=1em plus 2em minus .5em
\newdimen\digitwidth
\setbox0=\hbox{\rm0}
\digitwidth=\wd0
\catcode`@=\active
\def@{\kern\digitwidth}
\def\spose#1{\hbox to 0pt{#1\hss}}
\def\lta{\mathrel{\spose{\lower 3pt\hbox{$\mathchar"218$}}
 \raise 2.0pt\hbox{$\mathchar"13C$}}}
\def\gta{\mathrel{\spose{\lower 3pt\hbox{$\mathchar"218$}}
 \raise 2.0pt\hbox{$\mathchar"13E$}}}
\daectitle{Skewness induced by gravity}
\daecauthors{R. Juszkiewicz$^{1,2,3}$, F.R. Bouchet$^2$, \&
S. Colombi$^2$}
%
\daecinst{$^1$Copernicus Center, Bartycka 18, PL-00716 Warsaw, Poland}
\daecinsttwo{$^2$Institut d'Astrophysique de Paris, 98 bis Bd. Arago,
F-75014 Paris, France}
\daecinstthree{$^3$Institute for Advanced Study, Princeton, NJ 08540, USA}
\vskip 3cm
%
\daecabstract
Large-scale structures, observed today, are generally believed to
have grown from random, small-amplitude inhomogeneities,
present in the early Universe. We investigate how gravitational
instability drives the distribution of these fluctuations
away from the initial state, assumed to be Gaussian. Using second
order perturbation theory, we calculate the skewness factor,
$S_3 \equiv \langle \delta^3 \rangle \, /
\, \langle \delta^2 \rangle^2$. Here the
brackets, $\langle \ldots \rangle$,
denote an ensemble average, and $\delta$ is the density
contrast field, smoothed with a low pass spatial filter.
We show that $S_3$ decreases with
the slope of the fluctuation power spectrum;
it depends only weakly on $\Omega$, the cosmological density parameter.
We compare perturbative calculations
with N-body experiments and find excellent agreement over
a wide dynamic range.
If galaxies trace the mass, measurements of $S_3$
can be used to distinguish models with Gaussian initial conditions
from their non-Gaussian alternatives.
\vskip 3cm
\centerline{Accepted for publication in {\it Ap. J. Letters}}
\eject
\daecsec{1. Introduction}

Long before first estimates of the skewness in counts of
galaxies became available,
Peebles (1980) showed how gravity can generate skewness
in a random, initially Gaussian-distributed
density field in an Einstein-de Sitter cosmological model.
We have recently extended this calculation
to $\Omega \ne 1$ (Bouchet \etal 1992, hereafter Paper I).
In this {\it Letter}, we develop the formalism
further in order to bridge the gap between theoretical concepts
and observable quantities. We distinguish the mass density
contrast field, $\delta\rho/\rho$,
from the spatially smoothed field, $\delta$.
While the former is not directly observable, the latter may be.
Indeed, if the galaxies trace the mass,
the moments of the frequency distribution in the counts of galaxies
provide weighted averages of $\xi_m$ --
the $m$-point density correlation functions\footnote*{{\smallroman
We neglect the shot noise terms. All our calculations
are made in the fluid limit.}},
$$
\langle \delta^m \rangle = \int {dv_1\, \ldots \, dv_m \over v^m} \;
\xi_m ({\bf x}_1, \ldots , {\bf x}_m)\ ,
\eqno(1)
$$
rather than the proper moments, $\langle (\delta\rho/\rho)^m \rangle
= \xi_m(0)$. Here ${\bf x}_i$ are comoving spatial coordinates,
$dv_i \equiv F({\bf x}_i) d^3 x_i$,
while $F$ is a filter that determines the shape and volume,
$v \equiv \int d^3x\,F({\bf x})$,
of a resolution element.
%
In order to take the smoothing process into account, we calculate
moments of a weighted average of $\delta \rho ({\bf x})/\rho $,
$$
\delta ({\bf x}) \equiv \int {\delta\rho \over \rho}({\bf x'})
\,F({\bf x-x'})\,d^{3}x' \ .
\eqno(2)
$$
The filters considered here
are spherically symmetric, sweep a unit volume and
have a finite effective comoving half-width, $R$:
$$
\int F(x)\,d^3x = 1, \qquad{\rm and}\quad
\int F(x)\,x^2\,d^3x = R^2\ .
\eqno(3)
$$
We derive analytic expressions for the skewness of the
smoothed field $\delta$ in models with various
spectra of primeval fluctuations.
We consider two kinds of filters,
used by the observers and N-body simulators --
the Gaussian and the ``top hat'', or a sphere with a
fixed comoving radius. We show that the resulting
skewness is sensitive to the slope of the spectrum
of primeval fluctuations, as well as to the properties
of the filter, contrary to incorrect claims made
in the literature recently. We then test the limits of our approach
and compare perturbative results with N-body experiments.
We end by listing some still unresolved problems and
summarizing our results.
\eject
\daecsec{2. Perturbation theory}

Our calculations are based on the standard perturbative expansion
for $\delta\rho/\rho$  in a Friedman universe,
filled with non-relativistic pressureless
fluid and zero cosmological constant (cf. \S 18 in Peebles 1980).
We assume that to first order in perturbation theory,
$\delta\rho/\rho$ is a random Gaussian field. To first order, all its
statistical properties are therefore determined by the
power spectrum, $P(k) \equiv \int \xi_2 (x)\, \exp (i{\bf k \cdot x)}
\, d^3 k / (2\pi)^3$. Perturbative calculations also assume that
the amplitude of the density fluctuations, measured by their
variance, is small: $\langle \delta^2 \rangle \ll 1.$ The linear
order term in the perturbative expansion for $\langle \delta^2 \rangle$
is
$$
\sigma^2 \equiv \int\, {d^3 k\over(2\pi)^3} \,P(k)\, W_k^2  \; ,
\eqno(4)
$$
where $W_k = \int F(x)\,\exp(i{\bf k \cdot x})
\,d^3 x$ is the smoothing window function.
We consider three windows:
$W_k \equiv 1 \,$ (no filtering at all),
$W_k = (3/kR) \, j_1(kR) \,$, a top hat,
and $W_k = \exp (-k^2R^2/2) \,$, a Gaussian.
Here and below $j_{\ell}$ denotes a spherical Bessel function
of order $\ell$.
Deviations from Gaussian behaviour, induced
by gravity, appear at the second and higher orders and are fully
determined by $P(k)$. We define the {\it skewness factor}
as the ratio of skewness to variance squared,
$
S_3 \equiv \langle\delta^3\rangle\,\langle\delta^2\rangle^{-2}\ \, .
$
To lowest non-vanishing order, $S_3$ is given by
(Juszkiewicz \& Bouchet 1991)
$$
S_3 =
\int{d^3k\,d^3k'\over(2\pi)^6\,\sigma^4}
\;P(k)\,P(k')\,W_k\,W_{k'}\,W_{|{\bf k-k'}|}\,T({\bf k,k'})
+ O(\sigma^2) \; ,
\eqno(5)
$$
and the function $T({\bf k,k'})$ is given by
$$
T({\bf k,k'}) = 4 + 4\kappa(\Omega) -
6 \mu (k/k') + \left [ 2 - 4\kappa(\Omega)\right ] \, P_2(\mu)\ ,
\eqno(6)
$$
where $\mu = {\bf k \cdot k'}/kk'$ , and $P_2$
is a Legendre polynomial, while
$\kappa$ is a slowly varying function of
the current value of $\Omega$.
For densities in the range $0.05 \leq \Omega \leq 3$,
$$
\kappa(\Omega) \approx (3/14)\, \Omega^{-2/63} \; ,
\eqno(7)
$$
(Paper I).
The expression for $\langle\delta^3\rangle$ in ${\bf k}$
space, with $\Omega =1$, but without smoothing, was first derived by
Fry (1984), in a truly seminal paper.
Goroff \etal (1986), whose analysis was also
restricted to the $\Omega = 1$ case, were first to include the
filters $W_k$. In the absence of filtering
($W_k \equiv 1$), the dipole and quadrupole
terms in equation (6) integrate to zero, and we obtain
$$
{\textstyle
S_3 = 4 + 4\kappa(\Omega)
\approx {34\over7} + {6\over7}(\Omega^{-2/63}
- 1)}\ ,
\eqno(8)
$$
where the approximate form applies for $0.05 \le \Omega \le 3$.
The first term on the right hand side, 34/7, reproduces the
the Peebles (1980) result for $\Omega =1$. The second term, found
by Bouchet \etal (1992) is the ``curvature correction'',
which arises when $\Omega \ne 1$. This term is always small,
and $S_3$ is essentially insensitive to $\Omega$ (this was
pointed out independently by Martel \& Freudling 1991).
To second
order, the growth rates of the variance and skewness
are such that for any
comoving smoothing scale, the ratio $\langle
\delta^3\rangle \, / \, \langle \delta^2 \rangle^2$ remains
constant,  if $\Omega = 1$. In an open universe, $S_3$
grows extremely slowly as $\Omega$ decreases with time.
In the $\Omega \rightarrow 0$ limit, $\kappa \rightarrow 1/4$ (Paper I),
and $S_3 \rightarrow 5$, a tiny increase compared to $34/7 \approx 4.9$.
\par
Equation (8) can be regarded as an approximation,
valid in the regime when the density gradients
across the filtering scale $R$ are small.
Indeed, let us consider a density field
with a large coherence length,
$R_c \equiv \sigma/\sigma_1$,  where
$$
\sigma_1^2 = \int k^2\,P(k)\,W^2_k\,d^3k\,/(2\pi)^3
$$
is the variance of the density gradient.
In the limit $R\ll R_c$, we can Taylor
expand $W_k\,W_{k'}\,W_{|{\bf k-k'}|}$
about the origin, $k=k'=0$, and then evaluate the integral (5)
termwise. Using the normalization conditions (3),
and integrating over $\mu$, we obtain
$$
S_3 = 4 + 4\kappa(\Omega) - 2(R/R_c)^2
+ O(R/R_c)^4 \ .
\eqno(9)
$$
The first two terms above describe the ``local field''
contribution, as in eq.(8). The third term is the tidal correction.
Tidal effects, associated with the density gradients,
tend to lower $S_3$. This decrease is stronger for fields with
smaller coherence length. In the next section we will see that this effect
is also present for pure power-law spectra: $S_3$ {\it is anticorrelated
with the relative amount of small-scale power}.
\daecsec{3. Power-law clustering models}

We will now study power law spectra, $P \propto k^n$.
We set the smoothing length
to unity and use dimensionless wavenumbers.
Two kinds of filters are considered.
We begin with the top hat.
To separate
the variables $k$ and $k'$ in the multiple integral
(5), we use the addition theorem for Bessel functions
\footnote*{{\smallroman The formula for ${j_1}$ can be derived
by differentiating a similar expression for
$j_0$, given in Abramowitz \& Stegun (1964, eq. [10.1.45]);
for the full derivation, see Watson (1944).}},
$$
j_1(\omega) =  {\omega\over kk'}\,\sum_{\ell=1}^{\infty}
(2\ell+1)j_{\ell}(k)j_{\ell}(k'){dP_{\ell}\over d\mu} \; ,
\eqno(10)
$$
where $\omega = |{\bf k' - k}|$, and $P_{\ell}(\mu)$ are Legendre
polynomials of order $\ell$. After substituting the above
expansion to equation (5) and integrating over $\mu$, the
skewness factor can be expressed as
$$
S_3 = 4 + 4\kappa(\Omega) + \sum_{\ell=1}^{\infty}\,(6+4\ell)\,
{\beta(n,\ell)\over B^2 (n,0)}
\; ,
\eqno(11)
$$
where $\beta(n,\ell) = B^2(n,\ell)$ if $\ell$ is even,
and
$
\beta = - B(n+1,\ell)\, B(n-1,\ell)
$
otherwise.
The factors $B(n,\ell)$ are integrals of spherical Bessel functions,
$$
B(-m,\ell) \equiv \int_0^{\infty}
{j_1(z)\,j_{\ell+1}(z)\over z^m}\,dz
= {\pi\,m!\,({\ell-m+1\over 2})!\over
2^{m+1} \, ({\ell + m\over 2})! \, ({\ell + m + 3\over 2})! \,
({m-\ell\over 2})!} \; .
\eqno(12)
$$
For $-3 \le n < 1$, the series converges. Summing it requires
some knowledge of the properties of the gamma function
and a high threshold against boredom. The reward is the simple
final result,
$$
S_3 = 4 + 4\kappa(\Omega) - (3+n) \; .
\eqno(13)
$$
This expression is extremely weakly sensitive to $\Omega$: the
first two terms above are, in fact, identical to those on the
right-hand side of equation (8). The skewness parameter is,
however, strongly sensitive to the spectral slope, $n$.
$$\vbox{\halign{\hfil$#$\tabskip=1em plus1em
minus1em&\hfil#\hfil&\hfil$#$\hfil\tabskip=1em&$#$\hfil\tabskip=1em
plus1em minus1em&\hfil #\hfil&\hfil#\hfil\tabskip=0pt\cr
\multispan6{\hfil\bf TABLE 1:\ Scale-free initial conditions\hfil}\cr
\noalign{\bigskip\hrule\smallskip\hrule\smallskip}
n&$F(x)^a$&\multispan2{\hfil $S_3$ (perturbation
theory)\hfil}&$S_3$(N-body)&$\sigma$\cr
\noalign{\smallskip\hrule\smallskip}
-3&G&34/7&= 4.9&--&--\cr
-2&G&(4\pi + 9\sqrt 3)/7&= 4.0&$3.8\pm 0.2^e$&0.7\cr
-1&G&(20\pi - 12\sqrt 3)/7\sqrt 3&= 3.5&$3.6\pm 0.2^e$&0.7\cr
0&G&8(4\pi\sqrt 3 - 17)/7\sqrt 3&= 3.1&$3.0\pm 0.2^e$&0.7\cr
1&G&8(11\pi - 12\sqrt 3)/21\sqrt 3&= 3.0&--&--\cr
-1&T&(34/7) - 2&= 2.9&$2.9\pm 0.1^b$&0.6\cr
0&T&(34/7)-3&= 1.9&$1.8\pm 0.3^c$&0.7\cr
1&T&&\phantom{=~} 1.9^d&$1.9\pm 0.1^b$&0.7\cr
\noalign{\smallskip\hrule\smallskip}
\multispan3{$^a$\ {\smallroman G = Gaussian filter; T = top hat.\hfill}}\cr
\multispan3{$^b$\ {\smallroman White (1992).\hfill}}\cr
\multispan3{$^c$\ {\smallroman Bouchet \& Hernquist (1992).\hfill}}\cr
\multispan3{$^d$\ {\smallroman Corrected for finite grid efects.\hfill}}\cr
\multispan3{$^e$\ {\smallroman Weinberg \& Cole (1992).\hfill}}\cr
}}
$$
We now consider the Gaussian filter, $W_k = \exp (-k^2/2)$.
The variance and skewness are finite for $n > -3$.
To evaluate the skewness integral (5) in this case,
we used a coordinate transformation which reduces
$({\bf k - k'})^2$ to a sum of squares.
Similar multiple integrals are considered by Rice (1954).
Our results for $\Omega = 1$ and several integer
values of the power index $n$ are listed in Table 1.
In Figure 1 we plot $S_3 (n)$ for $\Omega =1$ and
the entire range $-3\le n \le 1$, obtained
by computing the integral (5) numerically.
The case $n=1$ was considered earlier by Grinstein \& Wise (1984).
Again, when $\Omega \ne 1$, there is a small
additional ``curvature term'' in the expression
for $S_3$. We write this additional term as
$$
\Delta S_3\,(n, \Omega) \equiv
{\textstyle {6\over 7}}\,(\Omega^{-2/63} -1)\,p_n \; ,
\eqno(14)
$$
so that the Gaussian-smoothed skewness factor is
$S_3(n, \Omega) = S_3(n,1) + \Delta S_3$.
For $n = -3, -2, -1,$ and 0 we obtained $p_n = 1,\, 1.04,
\, 1.13,$ and 1.29, respectively.
For $n = -3$, both
$\langle\delta^3\rangle$ and $\langle\delta^2\rangle^2$
diverge. However, their ratio is finite and
equal to $4 + 4\kappa$.
Note that in all cases, discussed in this section,
the dependence of $S_3$ on the
relative amount of small-scale power is
qualitatively similar to that
in equation (9): $S_3$ is a decreasing function of $n$.
\daecfigure{1}{8.5}{{\smallroman The skewness factor for power law
spectra and Gaussian smoothing. Rigorous perturbation
theory agrees well with the N-body results, while
the Zel'dovich approximation systematically
underestimates $S_3$.}}
\daecsec{4. Comparison with N-body experiments}
In Table 1 we summarize the results of several
N-body experiments, used to study
the evolution of skewness in the weakly nonlinear
regime for scale-invariant initial conditions.
The first column gives the power index, $n$. The
second column specifies the spatial filter used in
the simulation. The third column contains
our predictions for $S_3$, based on the perturbation theory.
The fourth column gives $\sigma$,
measured at the same time and
on the same smoothing scale in the simulation, as $S_3$.
The examples $n = +1$ and $-1$ at the bottom of Table 1
were provided to us by Simon White,
who calculated $S_3$, using the output from
N-body simulations of Efstathiou \etal (1988).
Similar results for $S_3$ were recently obtained
by Bouchet \& Hernquist (1992) and
Lahav \etal (1993) in their simulations of later
stages of clustering ($\sigma \approx 1$). For a Gaussian
filter, we used the recent numerical results of Weinberg \&
Cole (1992; we quote sampling errors as estimated by David
Weinberg, private communication).
\par
We have also considered the cold dark matter model (regarded here
as no more than an example of a spectrum with a
scale-dependent slope). The skewness factor was calculated by numerical
integration of eq. (5) and compared to an N-body simulation.
For the simulation, we used the particle-in-cell code, described
in Moutarde \etal (1991). This simulation involves
$64^3$ particles on a $128^3$ grid. We assume the `standard'
spectrum with a Hubble constant
of 50 km s$^{-1}$ Mpc$^{-1}$, $\Omega = 1$, and no biasing.
In Table 2 we list results for several
smoothing widths $R$, and for two kinds of filters.
All measurements were made at the time,
when the r.m.s. fluctuation in the density field, smoothed with
an $R = 16$ Mpc top hat, reached 0.92.
$$\vbox{\halign{\hfil$#$\tabskip=1em plus 1em minus1em
&\hfil#\hfil&\hfil$#$\hfil&$#$\hfil&\hfil$#$\hfil\tabskip=0pt\cr
\multispan5{\hfil\bf TABLE 2:\ $
S_{\bf 3}$ for a variable slope spectrum
(CDM)\hfil}\cr
\noalign{\bigskip\hrule\smallskip\hrule\smallskip}
\hfil R^a\hfil&$F(x)^b$&S_3{\rm
(perturbative)}&\multispan1{\hfil $S_3$(N-body)\hfil}&\sigma(R)\cr
\noalign{\smallskip\hrule\smallskip}
5.76&G&\phantom {c}3.70 \pm 0.07^c&3.95 \pm 0.15^d&1.30\cr
8.12&G&3.60 \pm 0.07&3.70 \pm 0.20&0.92\cr
11.52&G&3.50 \pm 0.07&3.50 \pm 0.15&0.64\cr
16.28&G&3.40 \pm 0.07&3.30 \pm 0.10&0.44\cr
23.00&G&3.30 \pm 0.07&3.20 \pm 0.10&0.29\cr
32.60&G&3.25\pm 0.10&3.05\pm 0.10&0.18\cr
11.52&T&3.15 \pm 0.10&3.80 \pm 0.38&1.30\cr
23.00&T&2.75 \pm 0.10&3.05 \pm 0.30&0.64\cr
46.00&T&2.25 \pm 0.25&2.45 \pm 0.25&0.29\cr
65.20&T&2.00 \pm 0.50&2.20 \pm 0.20&0.18\cr
\noalign{\smallskip\hrule\smallskip}
\multispan3{$^a$\ {\smallroman Smoothing scale in Mpc.\hfill}}\cr
\multispan3{$^b$\ {\smallroman G = Gaussian filter; T = top hat.\hfill}}\cr
\multispan3{$^c$\ {\smallroman Monte-Carlo integration errors.\hfill}}\cr
\multispan3{$^d$\ {\smallroman Sampling errors.\hfill}}\cr
}}
$$
We were able to find only
two cases of seemingly serious disagreement
between perturbative predictions and numerical experiments.
\par
The first case is the $n=1$ and a top hat
filter, when the perturbative series (13) diverges:
$S_3 = \infty$. Meanwhile, N-body experiments
(Lahav \etal 1993, White 1992) give $S_3 \approx 2$.
This `discrepancy' is easy to understand. The divergence
is caused by the $P(k)\propto k$ behaviour at large wavenumbers.
Such a spectrum is not reproduced in simulations
for wavenumbers larger than the Nyquist frequency, $k_N$,
defined by the particle grid. To model this effect,
we replaced the scale-free $n = 1$ spectrum, with
$$
P(k) = \cases{A\, k,&if $k\le k_N$;\cr
              A\, k_N,&otherwise, \cr}
\eqno(15)
$$
where $A$ is a constant, and $k_N$
matches the grid used by Efstathiou \etal (1988).
After this modification, the skewness integral (5) converged to
$S_3 = 1.9$, in perfect agreement with the N-body
results (Table 1).
\par
The second case of seeming discrepancy is the $S_3$ obtained
by Park (1991) for $n = -1$ and a Gaussian filter.
The source of problem in this case is the Zel'dovich (1970) approximation
(hereafter ZA), used by Park instead of an N-body code
to calculate particle trajectories.
ZA does not conserve momentum at second order in perturbation
theory. This leads to an incorrect form for $T({\bf k,k'})$,
with $\kappa$ in equation (6) set to zero (Grinstein \& Wise 1986; Paper I).
Using this incorrect expression, we were able to reproduce
Park's result (Figure 1).
Clearly, ZA systematically underestimates $S_3$
and disagrees rather badly with the rigorous perturbation theory
and with the N-body results.

\daecsec{5. Discussion}

Our results, summarized in equation (14) and Table 1,
show that the skewness parameter is a decreasing
function of the slope of the power spectrum; it
is also sensitive to the shape of the window function used.
\par
Gravitationally induced skewness was also investigated by
Coles \& Frenk (1991, hereafter CF), who claim that $S_3 \approx 3$
is a universal constant.
The perturbative as well as N-body results, discussed here
do not support this claim. The cause of our differences with
CF is their failure to recognize that the perturbative
results they use are not universal but instead are spectrum-
and filter-specific. For example, their equation (11),
quoted from Grinstein \& Wise (1987, hereafter GW),
is valid {\it only} for
$n = 1$ and a Gaussian filter. More importantly, the
GW formula is inapplicable to the N-body simulations,
conducted by CF, because the spectra and filters
assumed do not match each other; for the same reason the
GW formula cannot be meaningfully compared with the
$S_3$, estimated from QDOT counts, considered by CF.
Despite our quantitative differences,
we do agree with CF on the qualitative level,
that a scale-independent $S_3$ can be a signature of
Gaussian initial conditions and a simple power-law
spectrum (see the last paragraph in this section).
\par
We compared our calculations
with results of N-body experiments
to see if the perturbative series, truncated
at second order, can lead to sensible results over
a broad enough dynamic range.
To our own surprise, the agreement between the
two sets of results remained excellent
even when the ``small parameter'' in the expansion,
$\sigma$, was close to unity. Apart from incorrectly
conducted or misinterpreted simulations,
all discrepancies appear
to be within the sampling errors of numerical experiments.
We also note that the agreement with N-body results
was systematically better for the Gaussian filter
than for the top hat, most likely because of high frequency
sidelobes, which made the top hat-smoothed $S_3$ sensitive to
strongly non-linear fluctuations at small scales.
\par
Remarkably, the qualitative properties of the weakly nonlinear
clustering, described above, appear to hold in the
nonlinear regime as well. When $\sigma \gg 1$,
at least for $\Omega =1$ and scale-free initial conditions,
the relation between $\langle \delta^3\rangle$ and
$\langle \delta^2\rangle$ is well described
by the semi-empirical formula
$$
\xi_3\,({\bf x}_1,{\bf x}_2,{\bf x}_3) = Q\,[\xi_2\,({\bf x}_1,
{\bf x}_2)\,\xi_2\,({\bf x}_3, {\bf x}_2)\, + \, {\rm sym.}\,] \; ,
\eqno(16)
$$
where $Q = Q(n)$ is a parameter, dependent on the initial spectral
index only (Peebles 1980; Efstathiou \etal 1988).
Adopting the {\it Ansatz} (16) is equivalent to setting
$T({\bf k,k'}) = 3\,Q$ in the skewness integral (5).
For a top hat filter, methods described in \S 3 yield
$$
S_3 \approx 3\,Q \,  + \, {\nu^2 \, (3+\nu)^2 \,Q\over
(5-\nu)^2\,(2-\nu)^4\,(3-\nu)^4} \; ,
\eqno(17)
$$
where $\nu$ is the spectral index in the strongly-nonlinear
regime, related to the initial slope
by the so-called scaling solution,
$\nu = - 6\,/\,(n+5)$ (Peebles 1980; if $n \geq -3$,
then $0 \geq \nu \geq -3$). Substituting $Q(n)$, measured
in N-body experiments (Efstathiou \etal 1988), we get
$S_3 = 4.5 , 2.9$, and 2.3 for $n = -1, 0,$ and $+ 1$, respectively.
Similar results were recently
obtained in N-body experiments of Weinberg \& Cole (1992),
who directly measured $S_3(n)$ in the nonlinear regime, and
by Fry \etal (1993), who found that $Q(n)$ decreases with
$n$. To summarise: the scaling $\langle \delta^3 \rangle
\propto \sigma^4$ seems to hold equally well both for $\sigma \ll 1$
and $\sigma \gg 1$; moreover, in both regimes
$S_3$ is a decreasing function of $n$.
\par
Preliminary results from observations appear
to be consistent with a scale independent $S_3$
(Saunders \etal 1991; Coles \& Frenk 1991; Park 1991;
Bouchet \etal 1991, 1993; Gazta{\~n}aga 1992; Lahav \etal 1993)
This is exactly what is expected in
the standard gravitational instability picture
with Gaussian initial conditions and a simple power-law spectrum.
It is even possible that strongly non-Gaussian models,
which give $S_3$, diverging like $\sigma^{-1}$ instead
of being constant, can already be excluded (Silk \& Juszkiewicz 1991).
However, before reaching dramatic conclusions,
more work is needed. For example, the absence
of rich galaxy clusters in the IRAS survey is likely to cause
a systematic underestimate of $S_3$. We need
to understand how $S_3$ in the matter distribution relates
to the skewness in galaxy counts, as matter and galaxies may be
distributed differently. Finally, it is necessary to
account for the effect of redshift space distortion on $S_3$.
Of all of the unresolved
problems listed above, the latter admits the most
straightforward solution, and we plan to report on this
in near future.
\daecakno
RJ thanks David Spergel and Chris McKee
for raising his interest in deviations from Normal behaviour,
and Alain Omont, John Bahcall and Simon White
for their hospitality in Paris, Princeton and Cambridge, respectively.
We are grateful to Ofer Lahav, David Weinberg and Simon White
for sharing their unpublished numerical results with us.
We also acknowledge partial financial support from
the Polish Council for Scientific Research (KBN), grant
No. 2-1243-91-01. The computational means
(CRAY-2) were made available thanks to the scientific council
of the {\it Centre de Calcul Vectoriel pour la Recherche}.
\daecrefhead
\daecbook {Abramowitz, M. \& Stegun, I.A.}{1964}{Handbook of Mathematical
Functions}{Nationa Bureau of Standards}{Washington}

\daecproceed {Bouchet, F.R., Davis, M, \& Strauss, M.}{1991}
{``The Distribution
of Matter in the Universe'', 2nd DAEC Meeting}
{G.A. Mamon \& D. Gerbal}{Meudon}{Observatoire de Paris}{287}

\daecpaper {Bouchet, F.R. \& Hernquist, L.}{1992}{\apj}{400}{25}

\hangindent=3em\hangafter=1\smallroman
Bouchet, F.R., Juszkiewicz, R., Colombi, S. \& Pellat, R.,
1992, {\smallital Ap.J.,}{\bf 394}, L5 (Paper I).

\daecpreprint {Bouchet, F.R. {\smallital et al.}}{1993}

\daecpaper {Coles, P. \& Frenk, C.}{1991}{\mnras}{253}{727}

\daecpaper {Efstathiou, G., Frenk, C.S., White, S.D.M. \& Davis, M.}
{1988}{\mnras}{235}{715}

\daecpaper {Fry, J.N.}{1984}{\apj}{279}{499}

\daecpreprint {Fry, J.N., Melott, A.L. \& Shandarin, S.F.}{1993}

\daecpaper {Gazta{\~n}aga, E.}{1992}{\apj}{398}{L17}

\daecpaper {Goroff, M.H., Grinstein, B., Rey, S.-J. \& Wise,
M.B.}{1986}{\apj}{311}{6}

\daecpaper {Grinstein, B. \& Wise, M.B.}{1987}{\apj}{320}{448}

\daecproceed {Juszkiewicz, R. \& Bouchet, F.R.}{1991}{``The Distribution
of Matter in the Universe'', 2nd DAEC Meeting}
{G.A. Mamon \& D. Gerbal}{Meudon}{Observatoire de Paris}{301}

\daecpaper {Lahav, O., Itoh, M., Inagaki, S. \& Suto, Y.}{1993}
{\apj}{402}{387}

\daecpaper {Martel, H. \& Freudling, W.}{1991}{\apj}{37}{11}

\daecpaper {Moutarde, F., \etal}{1991}{\apj}{382}{377}

\daecpaper {Park, C.}{1991}{\apj}{382}{L59}

\hangindent=3em \hangafter=1\smallroman Peebles, P.J.E., 1980,
{\smallital ``The Large Scale Structure of the
Universe''\/}(Princeton: Princeton University Press).

\daecproceed {Rice, S.O.}{1954}{``Selected papers on noise and
stochastic processes''}{N. Wax}{New York}{Dover}{73}

\daecpaper{Saunders, W., {\smallital et al.}}{1991}{Nature}{349}{32}

\daecpaper {Silk, J. \& Juszkiewicz, R}{1991}{\nature}{353}{386}

\daecbook {Watson, G.N.}{1944}{A Treatise on the Theory of
Bessel Functions}{Cambridge}{Cambridge University Press}

\daecpaper{Weinberg, D.H. \& Cole, S.}{1992}{\mnras}{259}{652}
\hangindent=3em\hangafter=1\smallroman
White, S.D.M. (1992), private communication.

\daecpaper {Zel'dovich, Ya.B.}{1970}{\aa}{5}{84}
\bye